# The origin and influence of non-cavity modes in a micropillar Bragg microcavity

Matthew Jordan [1,2], Wolfgang Langbein [3], and Anthony J. Bennett [1,2*]
[1] School of Engineering, Cardiff University, Queen's Buildings, The Parade, Cardiff, CF24 3AA, United Kingdom
[2] Translational Research Hub, Maindy Road, Cardiff, CF24 4HQ, United Kingdom
[3] School of Physics and Astronomy, Cardiff University, Queen's Buildings, The Parade, Cardiff, CF24 3AA, United Kingdom
* Correspondence to: BennettA19@cardiff.ac.uk

**ABSTRACT**

Controlling the photonic environment of emitters is essential to the design of classical and quantum light sources. We study the case of a dipole-like emitter in a cylindrical pillar etched into a planar Bragg microcavity, which is a common design of quantum-dot single photon source. In addition to the well-known cavity modes created by the high-reflectivity of the Bragg mirrors at small in-plane wavevectors, we show the presence of broad spectral features that play a key role in controlling photon collection efficiency and Purcell enhancement. These 'non-cavity' modes are insensitive to the periodic index modulation of the Bragg reflectors, but arise from the cylindrical pillar geometry, as we show by comparison with simulations of uniform pillars, which reproduce the non-cavity modes. This approach provides a tool for understanding and modelling these often-disregarded decay channels as a function of source height, cavity dimensions and surface layers.

**I. INTRODUCTION**

The emission dynamics and efficiency of point sources in cavities is of both fundamental interest and critical for the development of photon sources for quantum technology. Cavity designs based around periodic variations in dielectric can result in high-quality factor modes, such as in photonic crystals [1-3], Bragg-cavities [4-9], and 'bullseye' circular gratings [10-12]. To achieve maximum Purcell factor and efficiency, emitters in these structures are generally assumed to couple maximally to the localised cavity modes.

The semiconductor quantum dot (QD) micropillar cavity is a popular choice of solid state device for the generation of indistinguishable single photons [13-15], as required by a wide range of quantum applications [16-19], courtesy of its ease of fabrication and high performance. These structures consist of a cavity spacer layer sandwiched between two distributed Bragg reflectors (DBRs). Reducing the lateral dimensions results in discrete optical cavity modes, localized between the mirrors, the lowest energy of which ($HE_{11}$) couples well to a far field collection optic such as a fibre or a lens [20,21]. Many studies have focused on maximizing the quality factor, $Q$ of $HE_{11}$ to obtain the highest photon collection efficiency [22,23] but this places a tight constraint on the detuning between the emitter's transition(s) and the mode, which can result in a low yield of efficient devices. Simulations have shown that even in an optimally coupled QD-cavity device the efficiency and Purcell factor display a periodic behaviour as a function of diameter [24,25]. This is a result of the interplay between coupling to the $HE_{11}$ and lower-$Q$ leaky "non-cavity" modes which radiate over a larger solid angle. It has recently been established that lower quality factor ($Q < 1000$) microcavities can also offer very high efficiency [24,26], still host dots capable of indistinguishable photon emission [15], and offer a more favourable tolerance to detuning. However, these lower-$Q$ devices are more strongly affected by the non-cavity modes, which justifies a detailed study of their origin and behaviour. These non-cavity micropillar modes are spectrally broad and overlap, making them difficult to extract from such simulations, but as they provide alternative decay channels for the source, understanding them is essential to maximize efficiency [27].

In the weak-coupling regime the emission into a single mode is parameterized by the Purcell factor $F_P$, defined as the ratio of total decay rate in the cavity - into the localized cavity mode $\Gamma_C$ and the other non-cavity modes $\Gamma_L$ [28] - relative to the decay rate in a homogenous medium $\Gamma_0$. In the case where $\Gamma_C \gg \Gamma_L$ the Purcell factor of an emitter of free-space wavelength $\lambda$ can be approximated with the mode volume $V$, quality factor $Q$, and effective refractive index $n_{\text{eff}}$ [29]:

$$F_P = \frac{\Gamma_C + \Gamma_L}{\Gamma_0} \approx \frac{3Q\lambda^3}{4\pi^2 V n_{\text{eff}}^3} \qquad (1)$$

Whereas the cavity modes are localized between the mirrors and can be simulated in finite-difference time-domain (FDTD) or finite element method (FEM) simulations using established numerical techniques, the leaky channels are often neglected from interpretation of the results. Although it is commonly assumed that $\Gamma_L \approx \Gamma_0$, this is an over-simplification [25], which neglects the influence of the non-cavity modes on the local photon density of states [30].

Here we study the relationship between the non-cavity modes and the gross shape of the adjacent semiconductor/air interfaces, by comparing simulations of cavities and identically shaped devices with a uniform refractive index. Simulations performed as a function of pillar diameter, source position and the position of the top surface are compared for cavities and uniform pillars. This work offers a method of modelling non-cavity mode behaviour, providing insights into their origin and potential optimization in future solid-state quantum light sources.

## II. RESONANT BEHAVIOUR OF NON-CAVITY MODES IN A BRAGG MICROPILLAR AS A FUNCTION OF DIAMETER

For all simulation results shown we use ANSYS Lumerical FDTD. We consider here a cylindrical micropillar consisting of a 267.9 nm $1\lambda$ GaAs spacer layer between with DBRs of $\lambda/4$ alternating $Al_{0.95}Ga_{0.05}As$ (78.9 nm) and GaAs (67.0 nm) layers, 7 above and 26 pairs below, on a planar GaAs substrate. This semiconductor stack is etched into cylindrical micropillars with vertical sidewalls running down to the substrate surface, with diameters between 1.5 and 4.0 μm. Material properties are modelled using the wavelength-dependent refractive index data for GaAs (3.55 at 940 nm) and $Al_{0.95}Ga_{0.05}As$ (2.94 at 940 nm). The refractive indices of the materials include dispersion and neglect absorption [31]. QD emission is modelled as an electric dipole source at $x = y = z = 0$, in the geometric centre of the spacer layer, oriented along the $x$-axis. $z$ is the symmetry axis of the cavity. The dipole is driven with a Gaussian pulse of 5.6 fs full width at half maximum (FWHM), covering a spectral range from 840 nm to 1070 nm. The simulation volume has $x$ and $y$ dimensions of 5.5 μm and a $z$ dimension of 10 μm centred on the dipole at the centre of the spacer, with perfectly matched layers at the simulation boundaries. Each surface of the simulation volume has a planar frequency-domain field monitor to record the local electric and magnetic fields and the Poynting vector, allowing the transmission spectrum to be calculated, through the surface of a cuboid around the pillar. Additionally, two planar frequency-domain monitors form cross-sections through the structure on the $x = 0$ and $y = 0$ planes. A schematic of the simulation environment, including these monitors, is shown in Fig. 1a.

The FDTD simulation outputs the spatial and spectral sum of all fields. Unlike complementary analytical simulation methods, such as the Fourier Modal Method [32,33] or Resonant State Expansion technique [30], FDTD does not explicitly separate different classes of mode. Thus, we define "cavity" modes as those arising from the strong reflection of the Bragg mirrors, which are spatially localized between said mirrors and have quality factors consistent with those mirrors' high reflectivity. All other spectral features are hereafter referred to as non-cavity modes. Fig. 1b shows the normalised far-field angular intensity distribution of the fundamental $HE_{11}$ cavity mode in a 2.00 μm diameter micropillar, as calculated from the electric field distribution at the top simulation surface monitor $T_{+z}$. Spectrally-resolved power through this monitor (Fig. 1c) and the complete surface formed by the six surrounding monitors (Fig. 1d), reveals a Lorentzian feature with a quality factor of ~600, consistent with analytical models of the $HE_{11}$ mode [34]. In addition to this, a broadband background of emission is observed, which dramatically modifies the spectrum at this micropillar diameter, indicating the importance of understanding these non-cavity modes.

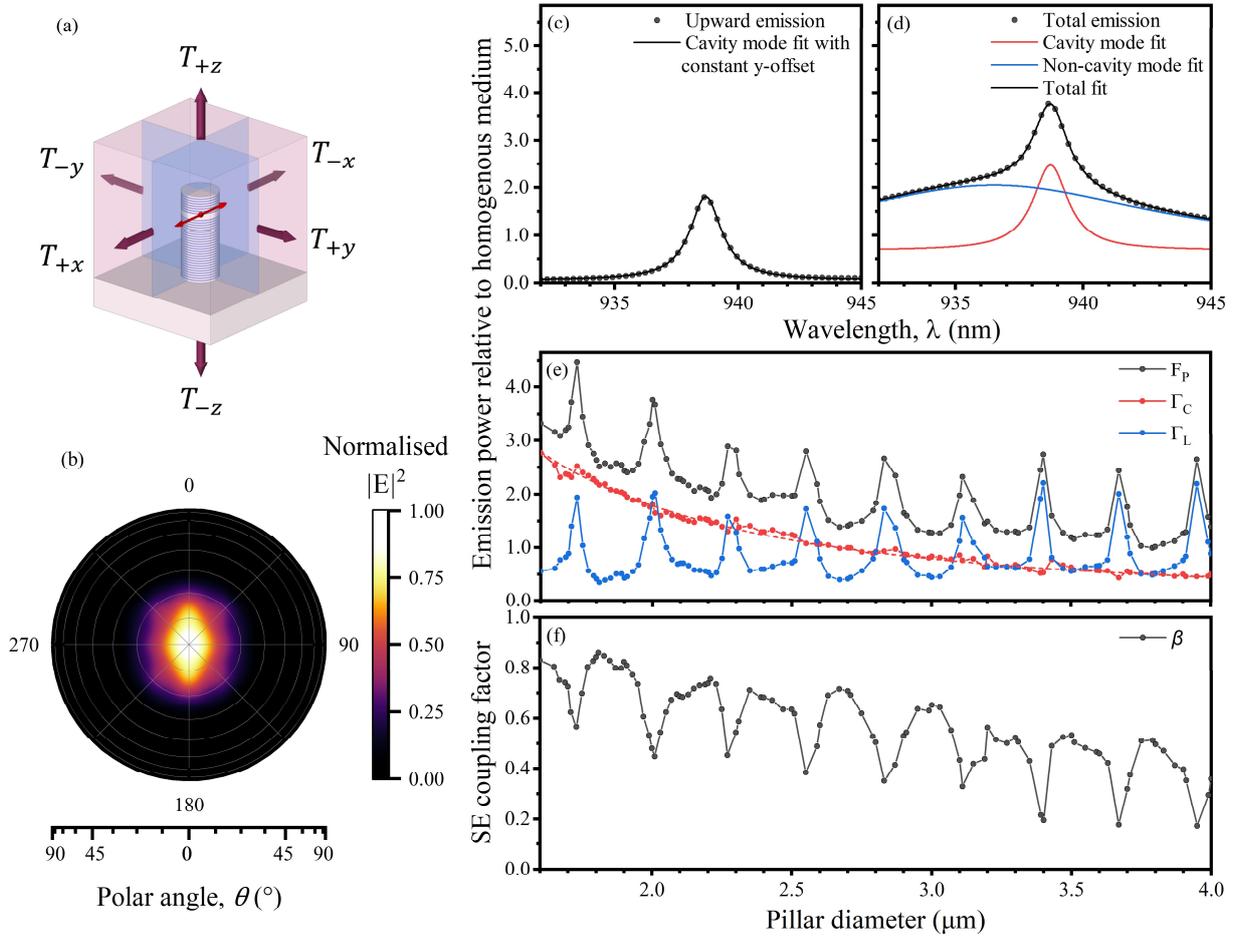

FIG. 1. Left: (a) Schematic of simulation environment, showing the micropillar on GaAs substrate, electric dipole source (red), cross-sectional (blue) and simulation boundary surface (purple) frequency-domain plane monitors, the latter labelled according to the transmission component measured through the monitor in each direction. (b) Simulated normalised far-field intensity, $|E|^2$, angular emission profile above a 2.00 μm diameter micropillar with 7 (26) upper (lower) DBR pairs. (c) Simulated emission power spectrum in the upward direction and (d) summed over all directions. cavity and non-cavity modes fitted with Lorentzian curves. (e) Purcell factor $F_P$, cavity mode emission $\Gamma_C$, and non-cavity mode emission $\Gamma_L$ and (f) spontaneous emission (SE) coupling factor $\beta$ as a function of pillar diameter. All emissions are normalised to the source emission power in a homogenous medium. The cavity mode emission rate was fitted with an inverse-square relationship as a function of diameter (dashed).

It is possible to extract the Purcell factor, cavity mode emission, and non-cavity mode emission at the $HE_{11}$ resonance wavelength from the total emission spectrum of the micropillar. First, the Purcell factor ($F_P$) will be equal to the total emission power normalised by the total power injected via the dipole at the cavity mode wavelength, which can be calculated as a sum of transmitted power through the closed surface of monitors. The proportion of emission via the cavity mode ($\Gamma_C$) on resonance can then be determined using a least-squares curve fit consisting of a Lorentzian cavity mode superimposed on a broadband Lorentzian background of non-cavity modes ($\Gamma_L$), as demonstrated in Fig. 1d. This serves as a fair approximation in this case, but it should be noted that there may generally be interference between modes, as they emit into non-orthogonal outgoing channels. The behaviour as a function of pillar diameter is plotted in FIG. 1e. The ratio of $\Gamma_C$ to the Purcell factor produces the spontaneous emission (SE) coupling factor $\beta$, which is plotted in Fig. 1f.

$\Gamma_C$ decreases as a function of diameter with an inverse-square relationship due to the increased mode volume (see equation (1)) [35]. Interestingly, $\Gamma_L$ varies periodically with diameter, having a period of 0.278 μm which is close to $\lambda/n_{GaAs}$, suggesting some resonant behaviour with the pillar sidewalls. $\Gamma_L$ takes values ranging from 0.35 to 2.24 $\Gamma_0$, showing a coupling to non-cavity modes ranging from suppression to enhancement, which is modulating $F_P$. This then also results in a periodic behaviour in $\beta$, which also decreases gradually with diameter due to the reduction in $\Gamma_C$. Although direct comparison of our results with previous literature is complicated by differences in design wavelength and number of repeats in the Bragg mirrors, other works have also reported periodic oscillations in $F_P$ and $\beta$ with diameter [23,25,26] arising from the interplay between cavity and non-cavity modes, and shown it has an important influence on the achieving the highest efficiencies and indistinguishability.

## III. COMPARING UNIFORM PILLARS WITH MICROPILLAR CAVITIES

To gain insight into the behaviour of the non-resonant modes without the effect of the resonant cavity, we replace the structured refractive index of the micropillar with a uniform index. We used the existing wavelength-dependent refractive index data for GaAs and $Al_{0.95}Ga_{0.05}As$ to determine a mean index weighted for the relative volumes of each material in the cavity (3.23 at 940nm). This averaging of the refractive index results in the same optical path length along the pillar.

Keeping the position of the electric dipole source, the resulting Purcell factor $F_P$ is given in Fig. 2 versus pillar diameter for both the Bragg cavity (a,b) and the uniform pillar (c,d). Panels (a,c) use only the emission transmitted through the top and bottom monitors, $T_{+z}$ and $T_{-z}$, which we call vertical emission, while (b,d) show the total emission spectra, including side or "lateral" emission, determined from the emission through all surfaces of the enclosed volume.

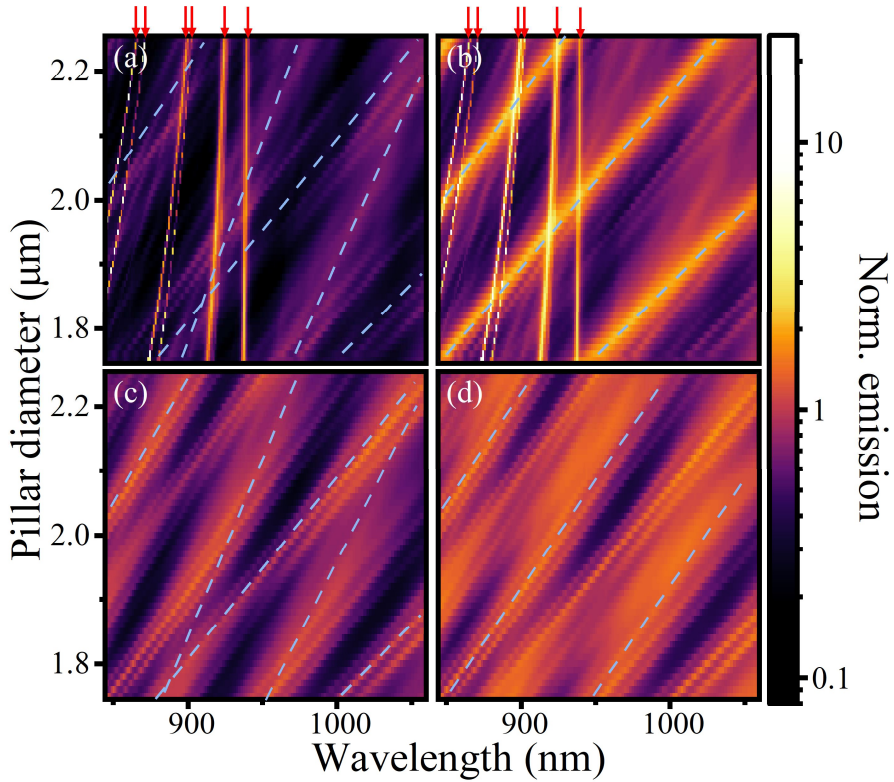

FIG. 2. Purcell factor for a source in the centre of the spacer layer, in a Bragg cavity (a,b) or uniform (c,d) pillar as a function of pillar diameter. In (a,c) only the emission through the top and bottom monitors $T_{+z}$ and $T_{-z}$ are considered while in (b,d) emission through all sides is considered. Red arrows indicate the positions of the cavity modes, whilst blue dashed lines highlight the gradients of some broad spectral features which appear in the cavity and the uniform pillar.

Disregarding the $HE_{11}$ and higher order cavity modes, visible as steep narrow streaks in (a,b) and marked with red arrows, which vanish in the absence of the cavity spacer layer, notable similarities exist between the two simulations. In particular, periodic broad diagonal features are visible which red-shift with diameter. In the vertical emission (a,c), two broad diagonal features cross one another with different gradients, indicated with blue dashed lines, as a result of different numbers of repeats in the top and bottom mirrors and thus the dipole source not being vertically centred in the pillar. The periodicities of the lines are well-reproduced by the uniform pillar, as are their phases, suggesting the uniform pillar is providing a useful approximation of the non-cavity modes. Previous work studying pillars of much smaller diameter using the Fourier Modal Method attributes broad spectral features to Bloch modes in the Bragg mirrors [32,33], which is supported by this and the following figure.

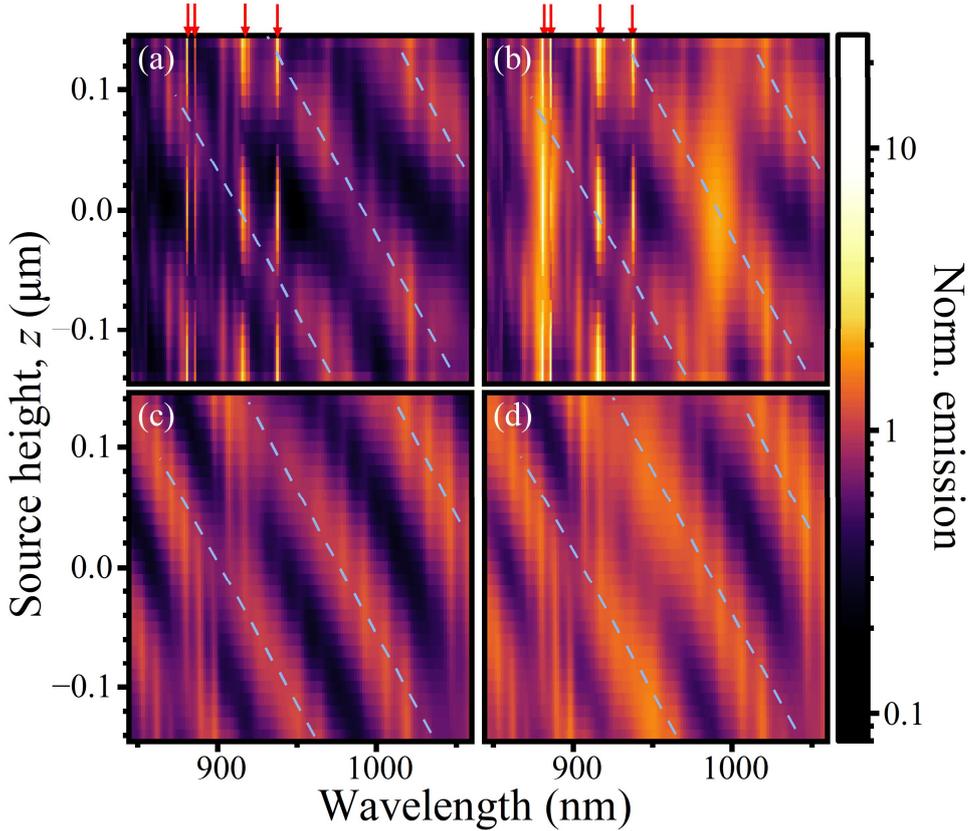

FIG. 3. Purcell factor for a source in a Bragg cavity (a,b) or uniform (c,d) pillar for a constant pillar diameter of 1.85 μm , as function of the source height. In (a,c) only the emission through the top and bottom monitors $T_{+z}$ and $T_{-z}$ are considered while in (b,d) emission through all sides is considered. Red arrows indicate the positions of the cavity modes, whilst blue dashed lines highlight the gradients of some broad spectral features which appear in the cavity and the uniform pillar.

This insight is further explored by changing the dipole source's height $z$ in pillars of 1.85 μm diameter, as shown in Fig. 3. The source height is varied over the full extent of the $1\lambda$ cavity spacer layer. In (a), the interaction between the source and cavity mode standing wave is apparent, having an anti-node in the centre and at the edges of the spacer. Beyond the sharp fundamental and higher order cavity modes, the broad features arising from the uniform pillar (c,d) display comparable broad features with the same periodicity and gradient to the non-cavity mode behaviour in (a,b). Non-cavity modes blue-shift with increasing $z$ at similar rates in both structures, and show similar periods, with a small offset in phase. This blueshift with dipole height is caused by the Bloch modes within the Bragg mirrors [23,32]: decreasing distance to the top surface, which provides a strong reflection shifts the resonance condition. The bottom surface to the substrate instead has only a small index step and accordingly has a lower reflectivity. The standing waves forming are thus similar to open pipe oscillations, which would suggest a scaling of the non-cavity mode's wavelength, $\lambda$, as a function of the distance between the emitter and the top surface of the pillar, $d$, varying as:

$$\lambda \propto \frac{1}{\sqrt{\frac{b^2}{r^2} + \frac{a^2}{d^2}}} \tag{2}$$

Where $r$ is the radius, and $a$ and $b$ are constants. Indeed, in both structures there are discrete bright resonances that pass-through nodes and antinodes as the dipole height is changed with a period greater than the wavelength, because we only see the z-component of the wavevector interact with the open-pipe modes along *z*.

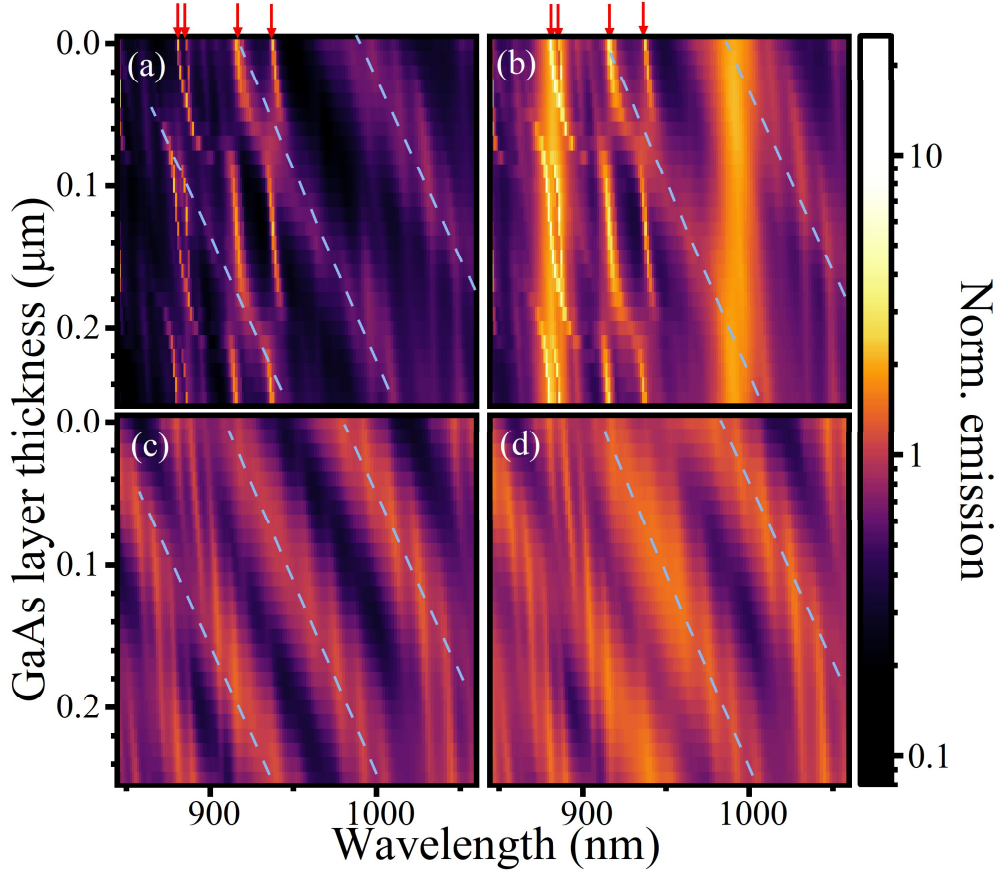

FIG. 4. Purcell factor for a source in the centre of the spacer layer, in a Bragg cavity (a,b) or uniform (c,d) pillar for a constant pillar diameter of 1.85 μm , as function of an additional GaAs layer added to the top of the pillar. In (a,c) only the emission through the top and bottom monitors $T_{+z}$ and $T_{-z}$ are considered while in (b,d) emission through all sides is considered. Red arrows indicate the positions of the cavity modes, whilst blue dashed lines highlight the gradients of some broad spectral features which appear in the cavity and the uniform pillar.

Finally, the pillars were modified by adding a layer of GaAs with varying thicknesses to the top surface (Fig. 4). This modifies the phase of the back-reflection from the top GaAs/air interface for the cavity mode, decreasing the Q factor of the cavity mode due to the change from constructive to destructive interference at $\lambda/4$ thickness of about 70 nm and back to constructive at $\lambda/2$ thickness. For the non-cavity modes, adding the layer is increasing the distance between emitter and surface ($d$). The resulting broad spectral features display similar periodicity and shift with similar gradients in the two models, as indicated by the dashed lines included in Fig. 4. This may be expected because they arise from either reflections of the Bloch modes in the top Bragg mirror reflecting from the cavity top surface or from the refection of the same surface in the uniform pillar. The distance from the dipole to the GaAs substrate does not change. Evidently, the addition of this GaAs spacer has implications for the optimisation of the pillar diameter and photon collection strategy.

## IV. FIELD PROFILES OF CAVITIES AND UNIFORM PILLARS

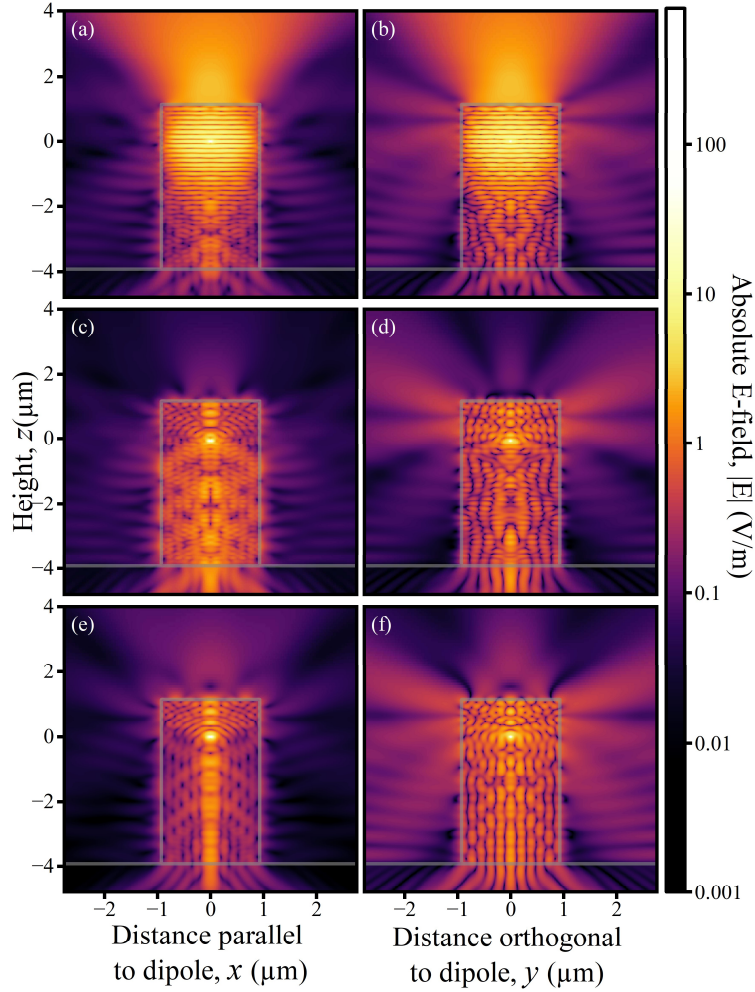

FIG. 5. Absolute electric field |E| cross-sectional profiles for 1.85 μm micropillars (grey outline) at the $HE_{11}$ mode wavelength for a cavity pillar excited by a dipole at the cavity mode anti-node ($z = 0$ μm) in the centre of the spacer layer, in (a) on the $y = 0$ plane ($x$-$z$), parallel to the dipole and in (b) on the $x = 0$ plane ($y$-$z$), perpendicular to it. (c, d) as (a, b), but for a dipole at the lower cavity mode node ($z = -0.067$ μm). (e,f) as (a, b), but for a uniform micropillar.

Overall, the model which uses a uniform micropillar consisting of a material with the weighted mean of GaAs and $Al_{0.95}Ga_{0.05}As$ refractive indices provides a good approximation to the non-cavity modes of a low $Q$ micropillar. The reason for this is elucidated by the cross-sectional electric field profiles in FIG. 5. Fig. 5: simulations of the cavity at the resonant wavelength and with a dipole at the field anti-node shows, as expected, the dominant coupling to the $HE_{11}$ mode. Emission is mostly directed into the $+z$ direction (a,b) due to the larger number of bottom Bragg pairs. However, when the source is moved to the $HE_{11}$ node it couples only to the non-cavity modes resulting in a complex field pattern (Fig. 5c,d). Similarities are visible in the field patterns within the uniform pillar when the dipole is at the same height (e,f). This supports the idea that the non-cavity modes have a similar spatial distribution to the uniform pillar which is to be expected when one considers the similarities between the Bloch modes in the Bragg mirrors [32,33] and the standing waves in the uniform pillar.

Notably, we see in (e) a beam in the centre of the pillar, with is dominated by modes with a radial quantum number of 5, counting the radial nodes in (f), showing the plane of strong dipole emission, orthogonal to the dipole. One can also note the directional emission towards the bottom, due to the reflection on the top surface.

## V. CONCLUSION

In conclusion, the non-cavity mode behaviour is reproduced by the uniform pillar with a refractive index taken to the be the weighted mean of the cavity indices. These non-cavity modes are dependent on the diameter of the micropillar, red-shifting much faster than cavity modes within the simulated wavelength range. Simulations where the emitter height is varied, and where an additional layer is introduced in both the cavity and the uniform pillar, confirm that the "non-cavity" modes are determined by the gross shape of the micropillar. Images of the field intensity in the uniform pillar also show similarities to the field intensity of an emitter at the field node in a cavity.

This insight can offer additional tuning knobs to intuitively design optimum cavities, where the aim is often to minimise the effect of the non-cavity modes. In future, uniform-index approximations of the pillar structure and analytical solutions for the non-cavity mode spectrum , such as that presented in equation (2), can narrow the parameter space required for exact FDTD optimisations. These insights are particularly relevant to low-$Q$ micropillars where the non-cavity modes have a greater influence on the performance metrics but are also relevant to efforts to engineer the most efficient high-$Q$ cavities.

## AUTHOR CONTRIBUTIONS

MJ: Conceptualization, Investigation, Data Curation, Writing - Original Draft, Writing – Review & Editing. WWL: Conceptualization, Writing - Review & Editing, Supervision AJB: Conceptualization, Writing - Review & Editing, Supervision, Project administration, Funding acquisition.

## DATA AVAILABILITY STATEMENT

The datasets used and/or analysed during the current study are available from the corresponding author on reasonable request.

## FUNDING DECLARATION

We acknowledge financial support from EPSRC Grant No. EP/T001062/1, EP/Z53318X/1 and EP/T017813/1.